\documentclass[10pt,conference]{IEEEtran}

\usepackage{color}
\usepackage{multirow}
\usepackage{pifont}
\usepackage{afterpage}
\usepackage{rotating}
\usepackage{subfigure}
\usepackage{color}
\usepackage{verbatim}
\usepackage{booktabs}
\usepackage[table]{xcolor}
\usepackage{listings}
\usepackage{arydshln}
\usepackage[neverdecrease]{paralist}
\usepackage{subfigure}
\usepackage{flushend}
\usepackage{hyperref}

\newcolumntype{L}[1]{>{\raggedright\let\newline\\\arraybackslash}p{#1}}  

\usepackage[
pass,
]{geometry}

\usepackage{url}
\urldef{\mailsa}\path|jguo@cs.mcgill.ca, natawutmo@gmail.com, janeclelandhuang@nd.edu|

\begin{document}
	
\title{Domain Knowledge Discovery \\Guided by Software Trace Links}
	
\author{\IEEEauthorblockN{Jin L.C. Guo}
		\IEEEauthorblockA{School of Computer Science\\McGill University\\Montreal, USA.\\
			jguo@cs.mcgill.ca}
		\and
		\IEEEauthorblockN{Natawut Monaikul}
		\IEEEauthorblockA{Dept. of Computer Science\\
			University of Illinois at Chicago\\
			Chicago, USA\\
			natawutmo@gmail.com}
		\and
		\IEEEauthorblockN{Jane Cleland-Huang}
		\IEEEauthorblockA{Computer Science and Engineering\\University of Notre Dame\\Notre Dame, USA.\\
			JaneClelandHuang@nd.edu}
		}

\maketitle
\begin{abstract}
Software-intensive projects are specified and modeled using domain terminology. Knowledge of the domain terminology is necessary for performing many Software Engineering tasks such as impact analysis, compliance verification, and safety certification. However, discovering domain terminology and reasoning about their interrelationships for highly technical software and system engineering domains is a complex task which requires significant domain expertise and human effort. In this paper, we present a novel approach for leveraging trace links in software intensive systems to guide the process of mining facts that contain domain knowledge. The trace links which drive our mining process, define relationships between artifacts such as regulations and requirements and enable a guided search through high-yield combinations of domain terms.  Our proof-of-concept evaluation shows that our approach aids in the discovery of domain facts even in highly complex technical domains. These domain facts can provide support for a variety of Software Engineering activities. As a use case, we demonstrate how the mined facts can facilitate the task of project Q\&A.
\end{abstract}

\section{Introduction}
\label{sec:Introduction}
Software and systems engineering projects are deployed across diverse domains such as communication and control, medical devices, finance, and electronic health care. Each of these domains is characterized by its own terminology \cite{DBLP:books/daglib/p/LuciaMOP12} which is used ubiquitously within relevant systems to specify requirements, create architectural documents, and to define variable names etc.  As a result, tasks such as identifying requirements ambiguities \cite{DBLP:journals/re/YangRGWN11}, evaluating requirements coverage in the design \cite{DBLP:journals/ese/HolbrookHDL13}, or performing a safety analysis \cite{DBLP:conf/icse/HatcliffWKCJ14} all require project stakeholders, and the tools they use, to understand the terminology of the application domain including the domain entities they represent and the relationships between them. 
The current Software Engineering practice of creating a basic project glossary defining terms and acronyms \cite{DBLP:conf/esem/AroraSBZ14,DBLP:conf/serp/ZouSC08} falls far short of providing the knowledge needed to automate tasks that are dependent upon natural language processing \cite{DBLP:conf/re/0004CB13}. Ideally, the glossary should be complemented with a full ontology defining domain specific terms and their concrete associations; however, constructing an ontology across highly technical software and systems engineering projects, can be extremely time-consuming \cite{DBLP:journals/kbs/GacituaSR08}.

We therefore propose a novel approach for mining domain knowledge in the form of associations between domain terms, referred to as domain facts from now on, for \emph{lightweight} ontology generation. We leverage the trace links inherent to all safety-critical software projects. Such projects are required by certifying bodies to provide explicit trace links across artifacts such as hazards, applicable regulations, design, source code, and test cases. A trace link is defined in terms of a \emph{source} and \emph{target} artifact. For example, consider the following trace link between a requirement and design element:\vspace{6pt}\\
\noindent
\begin{tabular}{p{2.1cm}|p{5.8cm}}
	\textbf{Artifact }&\textbf{Artifact Text}\\ \hline
	\textbf{Design (D1)} (Source) &The event thread records all actions or events for later review or audit.\\ \hline
	\textbf{ Requirement (R1)} (Target) &Each log entry shall have a time stamp with its time of occurrence 
\end{tabular} \\

\noindent This trace link indicates that design element $D1$ helps satisfy requirement $R1$. It also implies that associations exist between specific terms (e.g., nouns, noun phrases or verbs) in the text of the artifacts. Our approach searches for valid associations between each pairwise combination of terms across the source and target artifact.  For example, we may discover the domain fact that \emph{log} is a synonym for \emph{record} (typed relation) and/or that an \emph{event} is associated with \emph{time stamp} (untyped relation). Some associations embedded across trace links may not be described in a typical domain corpus, and therefore would be overlooked by conventional relation extraction methods.


\begin{figure}[t]
	\centering
	\includegraphics[width=.35\textwidth]{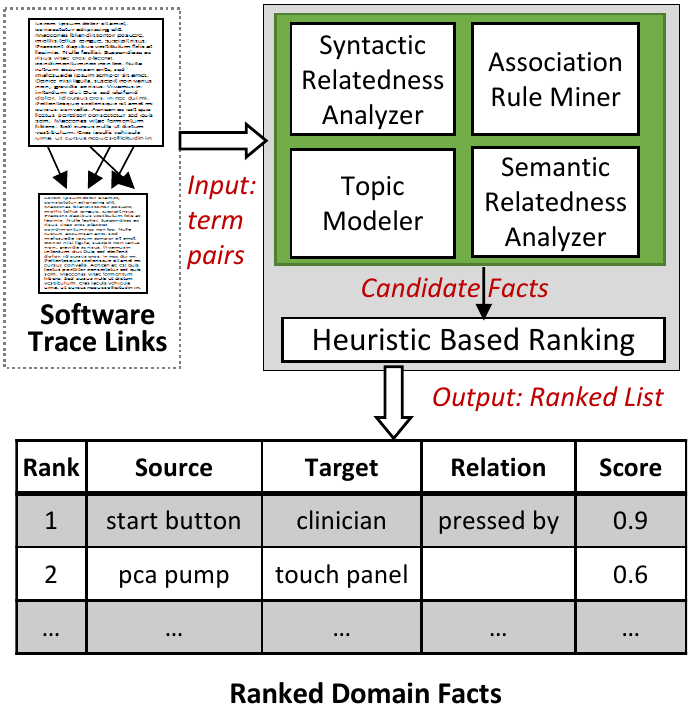}	
	\caption{ An overview of the relation mining process}
	\label{fig:Overview}
	\vspace{-10pt}
\end{figure}

We execute a \emph{targeted search} for domain facts guided by trace links.  Consider, for a minute an Electronic Health Record (EHR) system, which is one of the datasets explored later in this paper.  This EHR system contains approximately 13,000 domain specific terms. An unguided search for pairwise associations through this space would require investigating more than one hundred million combinations, the vast majority of which would be unlikely to yield useful associations. On the other hand, the project contains 1064 regulations and 116 requirements connected through 589 trace links. If, on average, each regulation and each requirement contains five domain specific terms, representing 25 possible associations per link, then our approach would search through only 14,725 potentially high-yield candidate associations. While it does not guarantee, and indeed is unlikely to find, a complete set of domain facts, it does find a subset of the most important ones inferred by existing trace links.  Most importantly, we have shown that even incomplete sets of domain facts can provide support for many software engineering tasks \cite{EMSEJin}.

We posit that the more complete grammatical structure inherent to software and system requirements facilitates the task of mining domain facts, and that such domain facts can then be used to support a broad range of tasks that depend upon natural language -- for example, understanding relations between domain terms may facilitate better support for activities such as issuing queries against source code.

\section{Domain Terminology}
\label{sec:DomainConcepts}

Our approach is illustrated and evaluated across the two domains of  \emph{Electronic Health Records} (EHR), and \emph{Medical Infusion Pumps} (MIP). The EHR project artifacts included 1064 regulatory requirements for managing electronic healthcare records, developed by the Certification Commission for Healthcare Information Technology (CCHIT) \cite{CCHIT}, 116 requirements extracted from documentation for WorldVista, the USA Veterans Health Care system \cite{WVista}, and 586 associated trace links between them. The MIP artifacts were all extracted from a Technical Report entitled ``Integrated Clinical Environment Patient-Controlled Analgesia Infusion Pump System Requirements" \cite{MIPRequirements}, which included goals, use cases, requirements, components, and trace links. We extracted 126 requirements, 22 component descriptions, and 131 associated trace links for purposes of this study. For each of the domains we also used Google to search for, and retrieve, 100 documents describing products in the domain. 

Those datasets contain 90,604 unique terms, i.e., words and phrases, in the EHR domain and 29,861 in the MIP domain. We differentiate domain-specific terms from more general ones by computing  \emph{domain specificity} as follows:
\begin{equation} 
\label{eq:DomainSpecificity}
DS(t)=ln \left[\frac{frq(t,D)}{\sum{_{t\in D}frq(t,D)}} \middle/\frac{frq(t,G)}{\sum{_{t\in G}frq(t,G)}}\right]
\end{equation}
\noindent where the first component $\frac{frq(t,D)}{\sum{_{t\in D}frq(t,D)}}$ is the normalized number of occurrences of term $t$ in the domain-specific corpus, and the second component is the normalized number of occurrences of $t$ in the general corpus of documents. Through experimentation we establish a threshold $T$, such that all words and phrases scoring DS scores higher than $T$ were deemed to be domain specific. 13,287 domain specific terms were found in the EHR system, and 4,317 in the MIP system.

\section{Mining Relations}
\label{sec:IdentifyCandidateFact}
The relation mining process is summarized in Fig. \ref{fig:Overview}. Given a trace link, it first evaluates the associations between each pair of domain specific terms extracted from the project's software artifacts using a variety of techniques. Candidate facts are generated and ranked by integrating those associations with a heuristic based method. The ranked list of candidate facts is then presented to the user for validation.

\begin{table*}[ht]
	\caption{Candidate Facts extracted using Lexico-Syntactic Patterns}
	\label{tab:Phrases}
	\centering
	\small\addtolength{\tabcolsep}{-3.5pt}
	\begin{tabular}{|L{1.4cm}|L{6.6cm}|L{5.6cm}|L{3.4cm}|}
		\hline
		\textbf{Type}&{\bf Regular Expression}&\textbf{Sample Sentence}&\textbf{Candidate Fact}\\ \hline
		{\bf Sub-Class} & $\ (and|or)\ (similar|other)\ $& ``Infusion pumps are used in \emph{hospitals} \textbf{and other} \emph{healthcare settings} worldwide.''&hospital {\bf is-subclass-of} healthcare setting \\ 		\hline
		
		{\bf Super-Class} & $\ (such\ as\ |including\ |eg\ |ie\ |(?<!that\ )(?<!to\ )include(s)?\ )$& ``\dots focused on how to compute ROI for \emph{medical IT systems} \textbf{such as} \emph{EHRs}.''&Medical IT System {\bf is-superclass-of} EHR\\ 
		\hline

		{\bf Is-Part-Of} &$\ (?<!that\ )(?<!to\ )(is|are|,|(can|must|shall|$ $may|might)be)\ (located|situated|found|$ $ incorporated)\ ([io]n|at\ )$& ``\emph{Audio speaker} \textbf{is located on} the \emph{the rear of the pump}.''&audio speaker {\bf is-part-of} rear of the pump\\ 
		\hline
		{\bf Has-Part} &$\ (?<!that\ )(?<!to\ )((consist(s)?of\ |$ $incorporate(?!d)\ |(is|are|,|(can|shall|must|$ $may|might)\ be)\ (made up|comprised) $ $of|contain(s)?\ ))$& ``Each \emph{profile} {\bf contains} \emph{instrument configurations} appropriate for the care area.''&Profile {\bf has-part} instrument configuration\\ 
		\hline	
		
	\end{tabular}
\end{table*}

\begin{table}[ht]
	\caption{Example of Candidate Facts extracted using Grammatical Structure Analysis}
	\label{tab:GSA}
	\begin{tabular}{|L{1.4cm}|L{6.2cm}|}
		\hline
		\multirow{4}{*}{\parbox{1.4cm}{\raggedright Sentence from MIP Domain Document}}& The drug library thread {\bf stores} the drug library provided by the hospital pharmacy and {\bf retrieves} the drug record corresponding to the liquid drug {\bf loaded into} the drug reservoir.\\ \hline
		\multirow{5}{*}{\parbox{1.4cm}{\raggedright Domain Specific Concepts}}&drug library thread\\ \cline{2-2}
		&drug library \\ \cline{2-2}
		&drug record \\ \cline{2-2}
		&liquid drug \\ \cline{2-2}
		&drug reservoir \\ \hline
		\multirow{3}{*}{\parbox{1.4cm}{\raggedright Candidate Facts}}&Drug library thread {\bf stores} drug library\\ \cline{2-2}
		&Drug library thread {\bf retrieves} drug record\\ \cline{2-2}
		&Liquid drug {\bf is loaded into} drug reservoir\\ \hline
	\end{tabular}

\end{table}

\subsection{Searching Associations}
For each pairwise combination of terms generated from trace links, we first employ four different techniques to search for associations between each pairwise combination of domain specific terms in the linked artifacts. As our results will indicate, different techniques identify different kinds of associations at varying degrees of confidence.

\vspace{3mm}\noindent{$\bullet$ \textbf{Syntactic Relatedness (SYN)}} is discovered using \emph{Lexico-Syntactic Patterns} (LSPs) \cite{Hearst:1992:AAH:992133.992154} -- generalized linguistic structures or schemas that indicate semantic relationships presented in the text \cite{DBLP:conf/nips/SnowJN04}. For example, from ``medications, \textbf{such as} antibiotics, \dots'' we can infer that \textit{medication} is a hypernym of \textit{antibiotics}. After reviewing domain documents, we identified a set of LSPs to extract taxonomic and compositional relationships and summarize them in Table \ref{tab:Phrases}. We also used \emph{ Dependency Path Analysis} to extract other types of non-taxonomic relationships by extracting domain specific terms that are connected by a verb or by a verb and a prepositional modifier, such as \emph{downstream monitor {\bf detects} air-in-line embolism}. Examples are provided in Table \ref{tab:GSA}. Syntactic relatedness techniques explicitly produce semantically labeled associations.

\vspace{3mm}\noindent{$\bullet$ \textbf{Semantic Relatedness (SEM)}} between pairs of terms is inferred through using WordNet \cite {miller1995wordnet}. WordNet organizes nouns, verbs, adjectives and adverbs as a semantic network of interlinked synsets. Following Lin \cite{lin1998information}, we calculate the Information Content (IC) of each synset and the IC of their \emph{Least Common Subsumer}(LCS) and compute relatedness as two times the IC of the LCS divided by the sum of the IC of each input synset. Accurately disambiguating word sense in a sentence is quite challenging; therefore, given two input words, we compute the similarity measure between all possible word senses and accept the largest value as the semantic relatedness measure. For phrases, we calculate  (1) the semantic relatedness between the head word of the phrase, and (2) the average semantic relatedness of all words from one phrase to its most related word in the other phrase. Additional candidate facts generated by the Semantic Relatedness approach are shown in Table \ref{tab:SrExample} along with their Semantic Relatedness measures.

\begin{table}[t]
	\caption{Candidate Fact Examples Learned from Semantic Relatedness (SEM)}
	\label{tab:SrExample}
	\begin{tabular}{|L{0.6cm}|L{2.1cm}|L{2.5cm}|L{0.6cm}|L{0.6cm}|}
		\hline
		& Source  & Target & HW & AW \\ \hline
		\multirow{4}{*}{MIP} &audible indication&drug record& 0.93 & 0.47 \\ \cline{2-5} 
		&drug reservoir&drug container& 0.65 & 0.98 \\ \cline{2-5} 
		&battery energy&main supply& 0.50 & 0.39 \\ \cline{2-5} 
		&infusion rate&hard limit& 0.47 & 0.25 \\ \hline
		\multirow{4}{*}{EHR} &restriction&business rule& 0.94 & 0.80 \\ \cline{2-5} 
		&medication order&real-time order check& 0.80 & 0.66 \\ \cline{2-5} 
		&patient group&user class& 0.58 & 0.42 \\ \cline{2-5} 
		&customize&hospital work flow pattern& 0.56 & 0.35 \\ \hline
		\multicolumn{5}{l}{Semantic Relatedness: HeadWord(HW),  All words(AW)}
	\end{tabular}
	\vspace{-6pt}
\end{table}

\vspace{3mm}\noindent{$\bullet$ \textbf{Association Rule Mining (ARM)}} is used based on the hypothesis that pairs of terms which appear more often across trace links are likely to be associated. For example, in the MIP dataset, the term \emph{scanner} was found in many requirements that were traced back to regulatory codes containing the term \emph{read}.  We found similar relationships such as between \emph{ICE interface} and \emph{bus transaction} and between \emph{light-emit diode} and \emph{hardware failure}. These kinds of associations can be learned using association rule mining (ARM) \cite{AS94}. 
Our interest is in frequent item sets of size two, for which one element is in the source artifact of the link and the other one is in the target artifact. Our goal is to capture associations between domain terms when their occurrence is strongly correlated, even if their appearance is rare. We therefore used the cosine measure to identify interesting associations, because it is not influenced by low probability events or by the total number of transactions. This measure ranges from 0 to 1, and a value close to 1 suggests a strong correlation between two items. 
We show several candidate facts generated from Association Rule Mining and their corresponding cosine measures in Table \ref{tab:ArmExample}. 

\begin{table}[t]
\caption{Candidate Fact Examples learned from Association Rule Mining (ARM)}
\label{tab:ArmExample}
\begin{tabular}{|L{0.6cm}|L{2.2cm}|L{2.5cm}|L{1.5cm}|}
	\hline
             & Source  & Target & ARM Val.\\ \hline
	\multirow{4}{*}{MIP} &hardware fault indicator&illuminate& 1 \\ \cline{2-4} 
	&normal operation&functional safety architecture& 1 \\ \cline{2-4} 
	&scanner&drug container& 0.81 \\ \cline{2-4} 
	&fault detection&exception& 0.71 \\ \hline
	\multirow{4}{*}{EHR} &us eastern standard time&different format& 1 \\ \cline{2-4} 
	&resolved problem&status column&0.82\\ \cline{2-4} 
	&prescription&local pharmacy& 0.76 \\ \cline{2-4}
	&user login/logout&time out& 0.7 \\ \hline
\end{tabular}

	\vspace{-6pt}
\end{table}

\vspace{3mm}\noindent{$\bullet$ \textbf{Topic Modeling (TM)}} is adopted to evaluate the co-occurrence of pairs of terms across the collection of all domain document and project artifacts. We used Latent Dirichlet Allocation (LDA) \cite{blei2003latent}, which is a high-performing generative probabilistic model that represents text corpora as a set of topics implemented in the MALLET\footnote{http://mallet.cs.umass.edu/}
topic modeling package, configured to identify 50 topics of 20 most probable terms. We compute the association between a pair of terms as the \emph{cosine similarity} of their corresponding vectors in the topic space. Scores range from 0 to 1, where values close to 1 indicate that terms often appear together in the same topics. The top terms for several topics from the EHR and MIP domains are shown in Table \ref{tab:topicExample}. It is important to note that neither Association Rule Mining nor Topic Modeling have the ability to semantically type the relationships they discover.

\begin{table}[t]
\centering
	\caption{Sample Topics Generated using LDA}
	\label{tab:topicExample}
	\addtolength{\tabcolsep}{-3.5pt}
	\begin{tabular}{|L{0.8cm}|L{0.3cm}|L{6.8cm}|}
		
		\hline
		
		\multicolumn{2}{|l|}{Topic}& Contributing Words and Probabilities                                                                \\ \hline
		\multirow{5}{*}{MIP} & 0        & infusion(0.17), program(0.073), volume (0.045), primary(0.045), secondary(0.037), \ldots \\ \cline{2-3} 
		& 1        & dose(0.13), pca(0.089), pause(0.058), bolus(0.053), infusion(0.032), \ldots \\ \cline{2-3} 
		& 2        & model(0.13), pump(0.12), alari(0.11), system(0.11), rev(0.088), \ldots \\ \cline{2-3} 
		& 3        & device(0.058), infusion(0.039), medical(0.029), summit(0.027), aamus(0.024), \ldots \\ \cline{2-3} 
		& 4        & drug(0.21), library(0.11), concentration(0.039), hospital(0.030), unit(0.016), \ldots \\ \hline
		\multirow{5}{*}{EHR} & 0        & action(0.094), patient(0.059), allergy(0.036), update(0.025), dob(0.024), \ldots \\ \cline{2-3} 
		& 1        & medical(0.036), university(0.022), association(0.020), american(0.019), informatics(0.016), \ldots \\ \cline{2-3} 
		& 2        & health(0.059), department(0.043), veteran(0.032), information(0.030), service(0.024), \ldots \\ \cline{2-3} 
		& 3        & encryption(0.028), device(0.028), user(0.024), storage(0.024), system(0.020), \ldots \\ \cline{2-3} 
		& 4        & care(0.077), patient(0.071), nov(0.067), physician(0.052), provider(0.050), \ldots \\ \hline
	\end{tabular}
	\vspace{-12pt}

\end{table}

\subsection{Rank and Validate Candidate Facts}
Each trace link can yield a large number of term combinations, referred to as candidate facts (CFs). For \emph{proof-of-concept} purposes we developed the following heuristics which utilize the four techniques described earlier to filter and rank the CFs. Our heuristics are based on observations about the individual and combined accuracy of the four techniques applied against the sample trace links.  We assume the terminology that $Ev_{TX}$ means that technique \textit{TX} supports the CF. Furthermore, \textit{DS} is domain specificity from Equation \ref{eq:DomainSpecificity} introduced in Section \ref{sec:DomainConcepts}, $S_{term}$ and $T_{term}$ represent the set of terms in source and target artifacts respectively.  \vspace{6pt}\\
\noindent 1. Remove the CF if $\neg Ev_{TM} \lor \neg Ev_{SEM}$.\\
\noindent 2. Assign confidence scores \textit{Conf} according to the following heuristics:\vspace{2mm}\\
\includegraphics[width=0.7\linewidth]{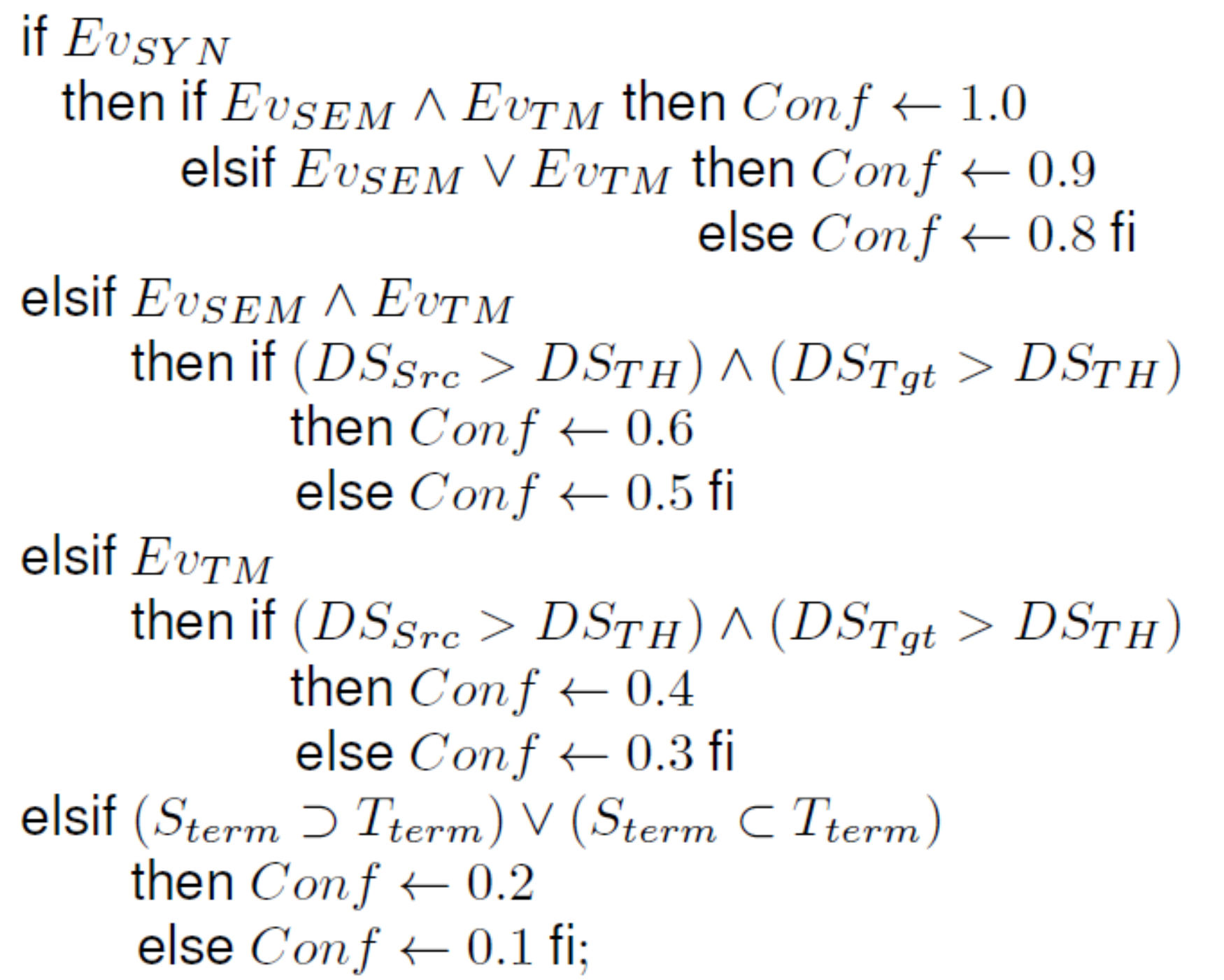}
\\
\noindent 3. Rank all CFs by \textit{Conf} score. In case of a tie, rank by the product of $Ev_{TM}$ and $Ev_{SEM}$ measures. For further tie, rank by the $Ev_{ARM}$ measure.\\
\noindent 4. Accept top \textit{N} candidate facts or all facts over a specified confidence score. 
\vspace{3mm}

At this point, the lightweight ontology can be generated in a fully automated way by accepting the top N candidate facts discovered from each trace link, or by accepting all facts over a certain confidence score.  Alternately, the ranked facts can be presented to a human analyst to allow them to vet the facts before saving them in the ontology. 

\section{Evaluation}
\label{sec:Evaluation}

 To evaluate the quality of the ranked candidate facts we designed and conducted an empirical study for both the EHR and MIP domains using 30 randomly selected trace links from each dataset.

 \subsection{Experiment Design} 
 For each of the two domains we recruited three participants who were asked to individually construct an \emph{answer set} of correct facts for each pair of linked artifacts. The participants were graduate students in Software Engineering at DePaul University. As domain experts were not available we provided the participants with several domain documents which they were encouraged to review and to reference in advance of, and during, the task of identifying facts. Each participant produced a list of facts associated with each of the linked artifacts. Any disagreement was resolved through discussion. The answer set ultimately represented a set of facts agreed upon by all three participants. For the EHR dataset, the answer set contained an average of 2.73 facts per link and a total of 80 unique facts for all 30 trace links. For the MIP dataset, it contained an average of 5.34 facts per link and 141 unique facts for all 30 links.

A ranked list of candidate facts was generated for each trace link using the techniques described in Section \ref{sec:IdentifyCandidateFact}. We then compared the generated facts to those found in the answer set.  The results are reported as a \emph{hit ratio} graph in which a fact generated by our approach and also found in the answer set is considered a hit. Graphs with steep initial increases are preferred because they represent cases in which correct facts are placed at the top of the ranked list. To investigate the effectiveness of individual techniques, we computed hit ratios for four individual techniques of semantic relatedness (using the `all term' variant), syntactic relatedness, association rule mining, and topic modeling.  We also present results achieved using our heuristic approach for integrating results from all four individual techniques. Finally, we draw the result achieved from randomly ranking the fact list as a baseline approach.

\subsection{Results and Analysis}
\begin{figure*}[h]
\centering
\begin{minipage}{.49\textwidth}
    \centering
    \includegraphics[height=2in]{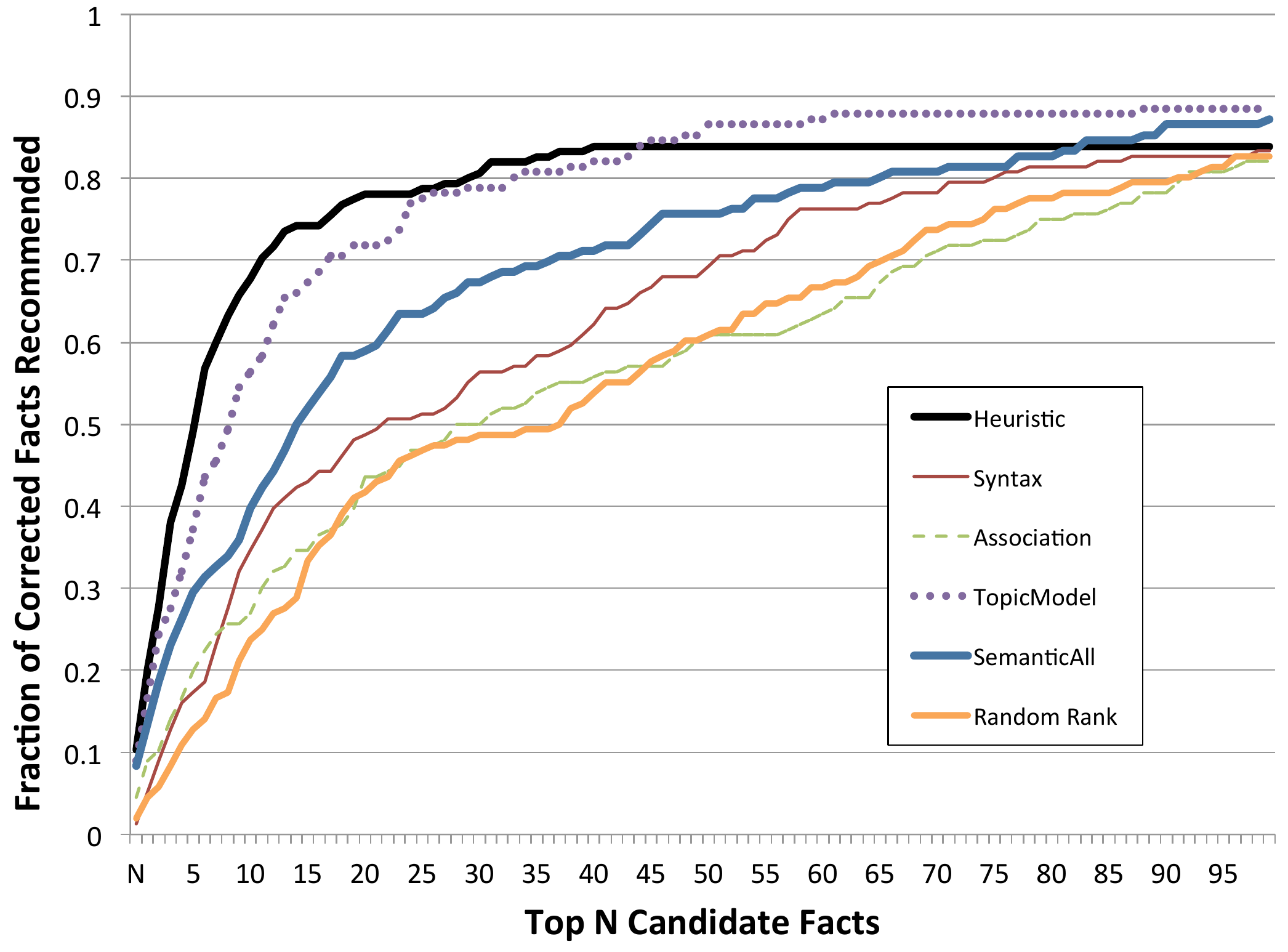}
\end{minipage}%
\begin{minipage}{.49\textwidth}
    \centering
    \includegraphics[height=2in]{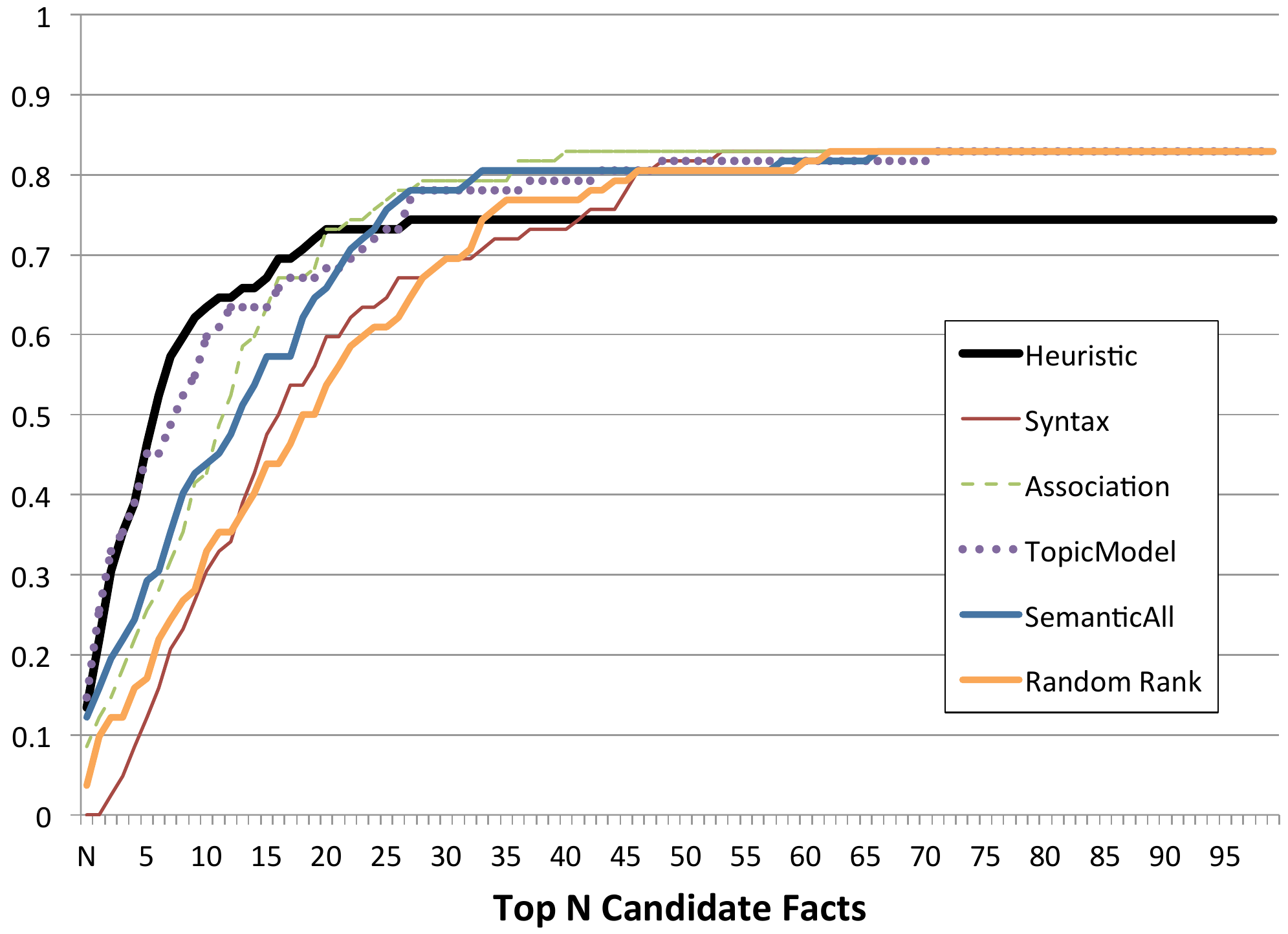}
\end{minipage}
\caption{Hit Ratio Graph for MIP Domain (left) and EHR Domain (right) showing Fraction of Correct Candidate Facts per top N Candidate Facts.}
\label{fig:HitRatio}
\vspace{-12pt}
\end{figure*}

The hit ratio graphs for each datasets are shown in Figure \ref{fig:HitRatio}. While we report results for completeness sake at values of $N$ ranging from 1 to 100, our interest lies primarily in values of $N$ in the range from 1 to 20 as it seems infeasible for an analyst to review more than 20 candidate facts per trace link. In fact, even lower values of $N$ seem more realistic.

In both domains we observe that the heuristic approach performed very well with approximately 50\% of the facts identified in the top 6 recommendations and approximately 75\% (EHR) and 80\% (MIP) identified in the top 20 recommendations.  The heuristic approach either outperforms or matches all other approaches. In EHR, similar performance at low values of $N$ were achieved using topic modeling. This technique also achieved good results for MIP indicating that topic models provide a relatively thorough coverage of the target domain. Association Rule Mining performs poorly in the MIP domain -- close to the baseline result.  This can be explained by the fact that there are only 131 trace links in the MIP  dataset, much fewer than the 589 links in the EHR dataset. Therefore, the MIP dataset offers less opportunity for learning meaningful association rules. Finally, neither the syntactic relatedness, nor the semantic analysis techniques performed very well in either datasets. Especially for the EHR domain, the syntactic relatedness technique only achieved similar results to the baseline. This is quite interesting, given that those are the techniques favored by the ontology learning community. However, we believe that this unexpected result can be explained by the unique environment of a software engineering project, and by the fact that the high performing topic modeling technique was constrained and guided by the traceability data.  This is a novel opportunity provided by using traceability data to mine domain knowledge. One important benefit of our integrated approach over an individual high-performing technique such as topic modeling, is that when multiple evidence sources include syntactic information, we are able to assign a label to the relationship.

The hit ratio graphs plateau between retrieving about 0.75 to 0.85 of the facts in both domains. There are three explanations for this.  First, certain facts were simply not retrieved using our approach because of thresholds we established within each of the techniques. Lowering these thresholds would negatively impact precision. Second, some errors occurred during the process of extracting phrases, especially when phrases were worded in unusual ways. Finally, a few facts were missed because they depended upon general phrases which were not included in our database of domain specific phrases. 

\section{Adding the Human in the Loop}
While our experimental results have shown that the automated approach discovered useful domain facts it can be beneficial to engage human analysts in the task of vetting and refining the results.  This is especially useful if the domain facts are going to be used across multiple projects.  To this end we developed a GUI tool which engages the user in the vetting process. The user may accept or reject the candidate facts and also modify the relationship type and even add or remove terms.  Table \ref{tab:RankedFact} depicts the facts mined using a trace link between a requirement and design element:\vspace{6pt}\\
\noindent
\begin{tabular}{p{2.1cm}|p{5.7cm}}
	
	\textbf{Artifact }&\textbf{Artifact Text}\\ \hline
	{\bf Design \newline (D1)} (Source) &The PCA pump shall have a start button.\\ \hline
	{\bf Requirement (R1)} (Target) &The control panel combines a touch panel with a speaker by which a clinician can enter and confirm configuration and see and hear alarms and warnings.\\ 	
\end{tabular} \\


\begin{table}[h]
\caption{Generated Candidate Facts }
\centering
\begin{tabular}{|L{0.5cm}|L{1.3cm}|L{1.5cm}|L{2.4cm}|L{0.7cm}|}
		\hline
		\textbf{Rank}& \textbf{Source Entity} & \textbf{Target Entity} & \textbf{Suggested Relation} & \textbf{Conf. Score} \\ \hline
		1&start button&clinician& press of $(Reverse)$& 0.9 \\ \hline 
		2&pca pump&touch panel&& 0.6 \\ \hline
		3&pca pump&control panel&& 0.5 \\ \hline 
		4&pca pump&alarm&& 0.5 \\ \hline
		5&pca pump&clinician&& 0.4 \\ \hline
		6&start button&touch panel&& 0.1 \\ \hline
		7&start button&control panel&& 0.1 \\ \hline 
\end{tabular}
\label{tab:RankedFact}
\end{table}
\section{Usage Examples}
\label{sec:Application}
Our purpose in building ontology is to support a variety of software engineering tasks -- including traceability, project level Q\&A, ambiguity analysis, and design satisfaction assessment.  As previously explained, all of these tasks could be performed better if a domain ontology were available.  In this section we illustrate the potential usefulness of a domain ontology for a specific SE task -- Project Q\&A.

\subsection{Project Level Q\&A}
\begin{figure}[!t]	
	\centering
	\includegraphics[width=.49\textwidth]{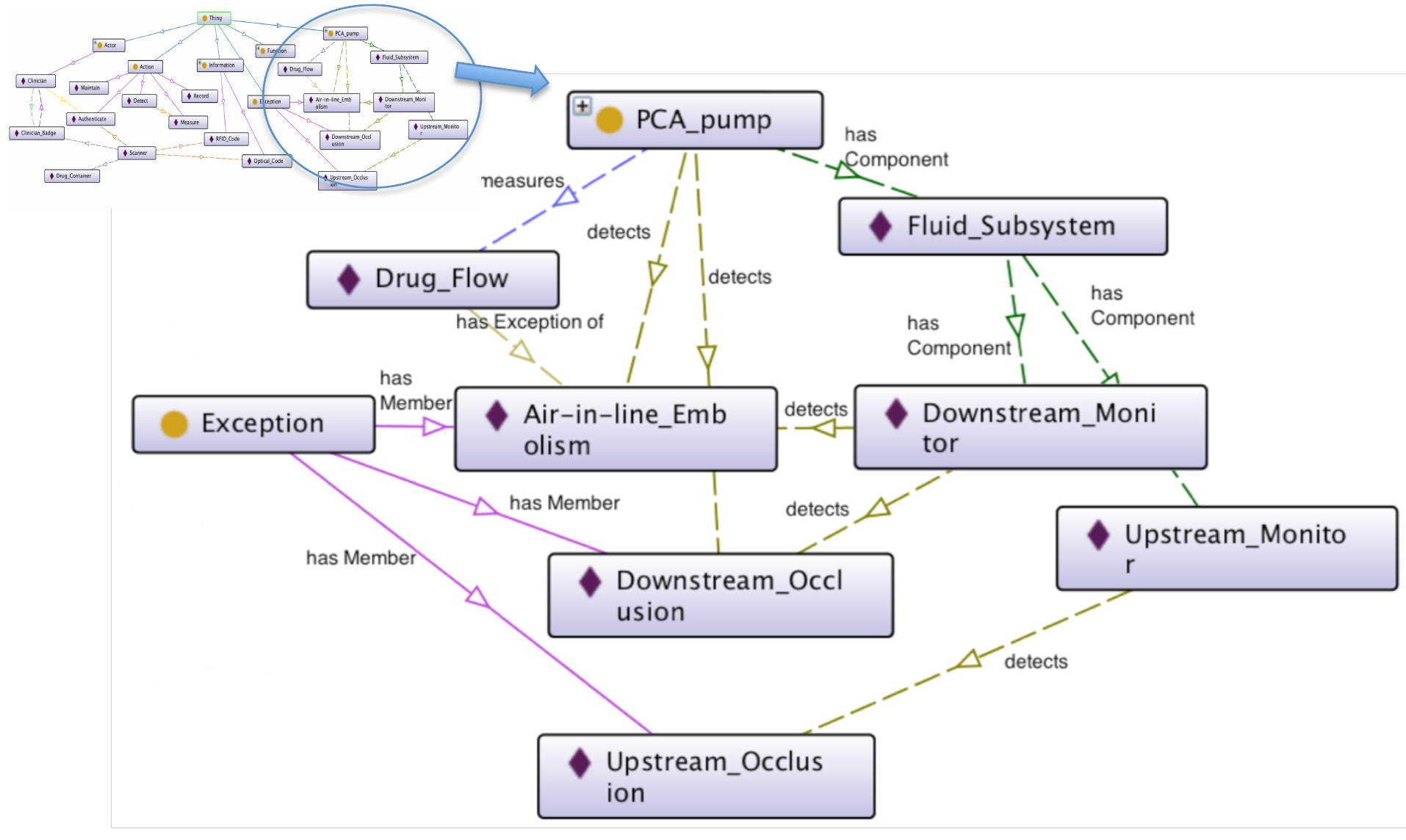}
	\vspace{-4pt}
	\caption{A subset of the MIP ontology after the human-in-the-loop process. }
	\label{fig:ProtegeView}
	\vspace{-8pt}
\end{figure}

In this example, we address the need that project stakeholders have for project intelligence. Software and Systems engineering projects accumulate a mass of data in the form of domain documents, requirements, safety analysis, design, code, test cases, simulations, version control data, fault logs, model checkers, project plans and so on. These artifacts are all specified using domain terminology. Existing tools such as \emph{TiQi}, which can transform natural language queries into executable queries \cite{DBLP:conf/re/PruskiLAOARC14}, could benefit from the presence of an underlying ontology. For example, a developer might ask TiQi to ``return all the requirements that demand the \emph{PCA pump} to catch fluid \emph{exceptions}''. It is not sufficient to just return classes containing the actual term \emph{PCA pump} and \emph{exceptions} because the query needs to be expanded to include other related concepts such as \emph{Downstream Monitor} (\textbf{subsystem} of PCA pump) and {Air-in-line Embolism} (\textbf{one type of} fluid exception). In Figure \ref{fig:ProtegeView} we provide the related part of the MIP ontology in which the relationships between these, and other terms, was discovered during our ontology mining process following the human vetting process.

\section{Threats to Validity}
\label{sec:Threats}
\textbf{External validity} evaluates the generalizability of the approach.  Our study included two different domains - with varying types and numbers of available software artifacts. Some of our findings are general in nature and apply to both of the studied domains.  For example, the accuracy rates of CFs for the top 6 recommendations were approximately 50\% in both domains.  However, other aspects differed across the domains.  For example, the average number of CFs that were relevant for each link varied greatly across the two domains, possibly because of differences in length and complexity of requirements.  Nevertheless, additional studies will be needed before claiming a general solution.

\textbf{Construct validity} evaluates the degree to which the claims were correctly measured. We evaluated the quality of generated candidate facts directly by comparing them to manually identified facts, as our purpose is to recommend valid domain facts that support the user's ontology building process. Other less direct evaluation techniques would fail to fully evaluate the accuracy of our approach. 

\textbf{Internal validity} reflects the extent to which a study minimizes systematic error or bias, so that a causal conclusion can be drawn. For both target domains, the answer set of domain facts was constructed by the three researchers.  We were not able to recruit domain experts for this task. To mitigate this problem we explained the target domain, defined what was meant by a correct fact to the participants beforehand, provided relevant domain documents, and encouraged participants to reference them throughout the study.  Finally, we consolidated the results of three participants, further reviewed reference materials, and discussed results with the participants until consensus was reached.

\section{Related Work}
\label{sec:Related}

\subsection{Ontology Construction}
There is a significant body of work in the area of ontology construction. Techniques can largely be classified as linguistic \cite{DBLP:journals/ao/Aussenac-GillesS05}, statistical \cite{DBLP:conf/swb/HeymansMAMSPMFKKSSFHWK08}, and machine learning \cite{DBLP:conf/nldb/Ruiz-CasadoAC05}. Initial efforts toward ontology building focused on matching lexico-syntactic patterns that occur repeatedly in free text \cite{Hearst:1992:AAH:992133.992154}. However, the recall using these methods is normally low due to limitations in defining sufficient patterns. Iterative bootstrapping techniques were introduced to overcome this problem. Starting from a small set of seed knowledge and patterns, these techniques discovered more patterns automatically and subsequently discovered new knowledge. To mitigate the ``semantic drift'' issue that occurs in bootstrapping techniques \cite{Curran07minimisingsemantic}, many strategies have been applied to constrain the learning process, such as type-checking relation arguments \cite{wang2012relation}. Researchers have also applied probabilistic models to evaluate and rank extracted relations \cite{Wu:2012:PPT:2213836.2213891}. More recently, researchers have focused efforts on mining ontology facts from large-scale public knowledge bases such as Wikipedia. Such knowledge sources typically include structured or semi-structured data, which is far easier to interpret in an automated fashion \cite{DBLP:journals/dke/Ruiz-CasadoAC07}. Unfortunately, for many software engineering domains, such as MIP and EHR, very limited structured information is available. Ontology mining therefore requires complex reasoning to extract and interpret important domain phrases and concepts.  Mining structured knowledge is therefore not directly applicable to the kinds of problems we are targeting in this paper.

\subsection{Ontology in Requirements Engineering}

There is a long history of ontology use in the area of Requirements Engineering, including presenting the requirements model itself, the acquisition structures for domain knowledge, the application domain, and the environment \cite{dobson2006revisiting}. 
Much work has been proposed to tackle the problem of ontology construction. For example, Breitman et al. described a bottom up ontology construction process that enables building application ontology during the requirements engineering activities \cite{OntolgyAsReProduct}; Kof proposed an approach to construct domain taxonomy by analyzing requirements
documents with natural language techniques \cite{kof2005appfication}; Omoronyia et al. extracted domain ontologies from technical documents specifically for supporting requirements elicitation task \cite{omoronyia2010domain}. Gacitua and Sawyer et al. proposed a framework for ontology construction \cite{DBLP:journals/kbs/GacituaSR08}, and explored different methods for assembling knowledge \cite{DBLP:conf/ACISicis/GacituaS08} with the primary focus on constructing the taxonomy of the target domain. 
The main difference in our work is the novel idea of leveraging traceability data to guide and constrain the association discovery process using a variety of techniques. In comparison previous techniques mainly used linguistic tools to analyze text from domain documents. Therefore, our approach can be used in conjunction with these solutions when traceability data is available to discover associations between concepts that do not explicitly appear in the domain or technical documents.

\section{Conclusion}
\label{sec:Conclusion}
In this paper we have presented a novel solution that leverages traceability data to guide and constrain the process of mining domain facts. The presence of a trace link can provide additional evidence which our experimental analysis has shown to effectively improve the process of generating domain facts.  

Our purpose in building ontology is to support a variety of software engineering tasks such as project level Q\&A, ambiguity analysis, and design satisfaction assessment. Our approach incurs the cost of ontology development in an initial project (or a set of initial projects), with expectation of realizing benefits across future projects.  For example, in a previous study we conducted \cite{EMSEJin}, we manually developed an ontology related to technical safeguards described in HIPAA (Healthcare Information Portability and Accountability Act) through analyzing the data in a single HIPAA-governed project.  We then demonstrated that the constructed  ontology could be leveraged to improve the quality of trace links automatically generated across six other unique HIPAA-related projects. The work presented in this paper introduces the feasibility of dynamically generating ontology for a specific domain, so that its benefits for supporting software engineering tasks can be realized across a far broader set of domains.

\section*{Acknowledgments}
The work in this paper was partially funded by the US National Science Foundation Grant CCF:1319680.

\vspace{10pt}
\bibliographystyle{abbrv}
\bibliography{Ontology}  

\end{document}